\documentclass[a4paper,fleqn]{cas-dc}

\usepackage{natbib}
\usepackage{pdflscape}
\usepackage{xcolor}

\usepackage{pdflscape}
\usepackage{adjustbox}
\usepackage{booktabs}

\definecolor{dgreen}{rgb}{0.0,0.45,0.0}
\newcommand{\bluebold}[1]{\textbf{\textcolor{blue}{#1}}}
\newcommand{\bluemag}[1]{\textbf{\textcolor{magenta}{#1}}}
\newcommand{\bluegreen}[1]{\textbf{\textcolor{dgreen}{#1}}}
\def\tsc#1{\csdef{#1}{\textsc{\lowercase{#1}}\xspace}}
\tsc{WGM}
\tsc{QE}

\begin{document}
\let\WriteBookmarks\relax
\def\floatpagepagefraction{1}
\def\textpagefraction{.001}

\shorttitle{}    

\shortauthors{}  

\title [mode = title]{A big step forward with SHARP: spatially resolved stellar population properties in passive galaxies at $z>1.5$}  

\tnotemark[1] 

\tnotetext[1]{This paper is based on a presentation discussing the SHARP science case for the Extremely Large Telescope (E-ELT).} 

%

\author[1]{A. Gargiulo}[orcid=0000-0002-3351-1216]



\ead{adriana.gargiulo@inaf.it}



\affiliation[1]{organization={INAF - Istituto di Astrofisica Spaziale e Fisica Cosmica (IASF-Milano)},
            addressline={Via A. Corti 12}, 
            city={Milano},
            postcode={I-20133}, 
            state={Italy},
            country={}}

\author[1]{C. Mancini}
\author[2]{F. R. Ditrani}
\affiliation[2]{organization={INAF - Osservatorio Astrofisico di Brera},
            addressline={via Brera 28}, 
            city={Firenze},
            postcode={I-20121}, 
            state={Italy},
            country={}}
\author[1]{S. Bisogni}
\author[1]{P. Franzetti} 
\author[1]{G. Vietri}           
\author[3]{A. Gallazzi}
\author[2]{M. Longhetti}
\author[3]{S. Zibetti}
\affiliation[2]{organization={INAF - Osservatorio Astrofisico di Arcetri},
            addressline={Largo Enrico Fermi 5}, 
            city={Firenze},
            postcode={I-50126}, 
            state={Italy},
            country={}}








\begin{abstract}
Understanding when and how massive quiescent galaxies ($\log(M_{\star}/M_{\odot}) > 10.5$) assembled their stellar mass and quenched remains a central challenge in galaxy evolution. 
Spatially resolved stellar population measurements at $z>1.5$ offer a uniquely powerful avenue to address this problem, as they can provide information on the radial variations in stellar age, metallicity, and enrichment histories in passive galaxies as they first emerge.
In this work, we present a feasibility study quantifying the transformative capabilities of the proposed IFU SHARP/VESPER at the ELT for performing such radial mapping of stellar population gradients in passive galaxies at $1.5<z<3$.
Using the COSMOS-Web catalogue, we define a realistic population of massive quiescent systems at $1.5<z<3$ and model representative compact and extended galaxies across this redshift range. Through detailed simulations with the official SHARP ETC, we derive the exposure times required to reach $S/N=10-15$ per resolution element at key rest-frame optical wavelengths. Our results show that SHARP will routinely measure stellar population gradients out to $2R_{\mathrm{e}}$ for the majority of the population at $z<2.5$ with integrations of $\sim20$\,h, and that will reach at least $R_{\mathrm{e}}$ in $\sim30$\,h at $z\sim3$. Thanks to MORFEO’s MCAO and to its spatial resolution of 30\,mas SHARP/VESPER will also resolve the inner $\leq1$ kpc at all redshifts considered, enabling for the first time, direct tests of quenching mechanisms linked to central mass build-up, bulge growth, and structural transformation.

These findings demonstrate that SHARP/VESPER will open an entirely new observational window on the early evolution of massive quiescent galaxies, providing, for the first time, statistically meaningful, spatially resolved stellar population constraints during the epoch when their stellar cores were assembled.
\nocite{*}

\end{abstract}





\maketitle

\section{Introduction}\label{}

\subsection{Formation pathways of massive passive galaxies}\label{sec:rationale_formation}

Understanding how massive quiescent galaxies assembled the bulk of their stellar mass and how their star formation was terminated remains one of the central challenges in galaxy evolution studies. A large fraction of today’s stellar mass resides in passive systems \citep{Renzini2006, Gallazzi08}, yet the physical processes that shaped their formation histories, chemical properties, and internal structure are still debated.

Over the past two decades, a wide range of observational studies spanning integrated spectroscopy, ground-based and {\it HST} photometry, converge on a consistent picture: massive quiescent galaxies assembled the bulk of their stellar mass early, typically at $z>1$–2. Direct evidence for this early formation has been provided by deep HST surveys, which revealed a substantial population of already quenched massive galaxies at these epochs \citep[e.g.][]{Daddi2005,Longhetti2007,Mancini2010,Saracco2010,vanderWel2014}.

JWST has now extended this evidence decisively to higher redshift. NIRCam and NIRSpec observations reveal a surprisingly mature passive population already established at $z\gtrsim 2$ \citep[e.g.][]{Suess2019,Suess2022,Suess2024,Carnall2023a,Carnall2023b,Wright2024}. Crucially, even at these early epochs, quiescent galaxies do not form a homogeneous class: at fixed stellar mass, their effective radii span more than a factor of five \citep[e.g.][]{Genin2025}, from extremely compact systems to more extended and lower density quenched galaxies. Such structural diversity strongly suggests that these galaxies did not follow a single evolutionary route. Some systems likely formed through rapid, dissipative events that built dense stellar cores, while others experienced more gradual or spatially extended assembly \citep[see e.g.][]{Gargiulo2017}. These distinct pathways were already imprinting a variety of mass assembly histories within the passive population at early cosmic times.

Despite their importance, integrated–light studies primarily constrain the global average properties of galaxies and provide only limited insight into the physical mechanisms that drove quenching and mass assembly. In contrast, spatially resolved analyses allow us to place stronger constraints on, for instance, whether bulges grew gradually or through rapid compaction, whether rejuvenation or accretion episodes occurred, and how the build-up of central regions relates to that of the outskirts. Part of this information becomes difficult to recover once spatial structure is collapsed into a single integrated spectrum.
Spatially resolved stellar population measurements are therefore essential, as they retain the radial signatures that directly reflect a galaxy’s formation and quenching history.

In the local Universe, integral-field spectroscopy has revealed that early-type galaxies typically display negative metallicity gradients, weak or flat age gradients, and nearly constant $[\alpha/{\rm Fe}]$ with radius \citep[e.g.][]{LaBarbera2012,Spolaor2010,Goddard2017,Gonzalez2015}. Results from the CALIFA survey further uncovered a characteristic ``U-shaped'' mass-weighted stellar age profile in many early-type systems \citep{Zibetti2020}, naturally interpreted within a two-phase assembly scenario: an early dissipative phase builds a dense, metal-rich core, while later ex-situ accretion deposits new stars at larger radii. These signatures place strong constraints on the balance between in-situ star formation and subsequent accretion-driven growth, as well as on the characteristic timescales of quenching. However, interpreting present-day gradients is complicated by the cumulative effects of dynamical evolution. Secular radial migration, repeated minor and major mergers, and even low-level rejuvenation episodes can significantly flatten or reshape the gradients that were originally imprinted at high redshift \citep[e.g.][]{Kobayashi2004,DiMatteo2009,Hirschmann2015}. As a consequence, local measurements alone provide only a blurred and temporally averaged view of the true formation pathways of massive passive galaxies.

\subsection{High-redshift resolved spectroscopy}\label{sec:rationale_highz}

Intermediate-redshift spectroscopic surveys have begun to bridge this gap. Programs such as LEGA-C \citep{Vanderwel2016} have provided high–signal-to-noise spectra for hundreds of galaxies at $0.6 \leq z \leq 1$, enabling the first statistically meaningful constraints on stellar population gradients beyond the local Universe \citep[e.g.][]{DEugenio2020,Cheng2024}. These studies confirm the persistence of negative metallicity gradients and generally flat age gradients, indicating that some elements of the internal structure of quiescent galaxies were already established by $z \sim 1$. 
However, by $z\sim1$, galaxies have already undergone several gigayears of evolution, including structural transformations, merger-driven mixing, and changes in their star-formation histories, all of which may partially erase the signatures imprinted during their early formation. Furthermore, datasets such as LEGA-C are based on VIMOS long-slit spectroscopy, obtained under typical VLT seeing conditions of $\sim$0.8". This seeing-limited setup imposes a fundamental constraint on spatial resolution, limiting the ability to robustly recover intrinsic stellar population gradients.

Attempts to infer stellar population gradients at high redshift have so far relied primarily on photometric measurements. Several studies have reported negative rest-frame colour gradients in quiescent galaxies at $1.5 < z \lesssim 2.5$, with redder centres and bluer outskirts, indicating the presence of underlying radial variations in their stellar populations even at these early times \citep[e.g.][]{Gargiulo2012,Chan2016,Ciocca2017}. The strength of these colour gradients appears to correlate with galaxy structure: compact or high–stellar-density systems tend to exhibit steeper gradients, whereas more extended quiescent galaxies show flatter or more heterogeneous trends \citep{Suess2019,Suess2022}. These behaviours suggest that the mechanisms responsible for quenching and early mass assembly are closely linked to the structural state of each galaxy at the moment star formation is suppressed. Yet colour gradients, by themselves, offer only a qualitative view of these processes: photometry cannot disentangle the strong degeneracy between age, metallicity, nor can it isolate the specific drivers of radial population variations. To move beyond these limitations and obtain physically robust constraints, spatially resolved spectroscopy is essential.

Measuring spatially resolved stellar population parameters from spectroscopic data requires access to the rest-frame optical absorption features that constrain stellar ages, metallicities, and elemental abundance ratios (see Longhetti et al., in this Science Book). Higher-order Balmer lines and the Balmer break trace stellar ages and recent star-formation histories, Fe-sensitive indices provide leverage on the overall metal content, and Mg-sensitive features probe $\alpha$-enhancement and the timescales of star formation. At $z>1.5$, these diagnostics fall in the near-infrared domain ($\lambda \sim 1$--$2.5\,\mu$m). In this regime, the combination of atmospheric transmission, strong sky emission, and, crucially, the intrinsic faintness of high-redshift passive galaxies demands instruments with substantially higher sensitivity and broad wavelength coverage. 

As a result, spatially resolved stellar population measurements beyond $z \sim 1.5$ have remained exceedingly scarce, limited primarily to studies of strongly lensed systems or low-resolution \textit{HST} grism spectroscopy \citep[e.g.][]{Ditrani2022,Akhshik2023,Jafariyazani2020}. Very recently, ultra-deep JWST/NIRSpec–MSA spectroscopy from the JWST–SUSPENSE program has revealed diverse and sometimes unexpected gradients in a sample of eight passive galaxies at $1.2 < z < 2.2$ \citep{Cheng2025}. While these results are groundbreaking, they rely on slit-based MSA observations at medium spectral resolution and modest spatial sampling, providing only sparse coverage of the central region (see Sec.\ref{sec:bigstep}). 

A transformative advance therefore requires an integral-field spectrograph operating in the near-infrared, with sufficient sensitivity to probe not only compact quiescent systems but also the intrinsically faint, low–surface-brightness outskirts of more extended galaxies, and with the spatial resolution needed to resolve the central kiloparsec where bulge growth, mass build-up, and quenching processes originate. These capabilities are crucial for mapping stellar population gradients in passive galaxies as they first emerge.

The SHARP spectrograph, equipped with the VESPER integral-field unit and assisted by MORFEO’s MCAO system, is designed precisely to meet this need. Mounted on the 39-m ELT, SHARP will benefit from the telescope’s unparalleled collecting area, which provides the sensitivity required to observe both compact quiescent galaxies and the intrinsically faint outskirts of more extended systems. MORFEO will deliver diffraction-limited spatial resolution of order 30\,mas in the near-infrared, corresponding to sub-kiloparsec scales at $z\sim2$–3. As outlined in the SHARP concept study (see Saracco et al. in this Science Book), VESPER will couple this spatial performance with a high spectral resolving power of $R \simeq 3000$ and a continuous wavelength coverage from 0.95 to 2.45\,$\mu$m. This unique combination will, for the first time, enable systematic and spatially resolved mapping of stellar population gradients during the epoch when quenching and early mass assembly occur.

Such measurements will directly test the different formation pathways proposed for massive quiescent galaxies. Rapid and highly dissipative events, such as gas-rich compaction phases or centrally concentrated starbursts, are expected to produce steep metallicity gradients and elevated $[\alpha/{\rm Fe}]$ ratios in the cores. In contrast, slower and more extended quenching processes driven, for example, by gradual gas depletion or environmental effects, should result in younger central ages and comparatively shallow abundance gradients \citep[e.g.][]{Kobayashi2004,DiMatteo2009,Hirschmann2015}. Identifying these distinct signatures at high redshift is crucial for reconstructing the mass assembly histories of passive galaxies.

We adopt a standard $\Lambda$CDM cosmology with $H_0 = 70\,{\rm km\,s^{-1}\,Mpc^{-1}}$, $\Omega_{m,0} = 0.3$, and $\Omega_{\Lambda,0} = 0.7$. All magnitudes are expressed in the AB system.

\section{The Sample}\label{sec:sample}

Assessing the feasibility of spatially resolved stellar population studies of passive galaxies at $z>1.5$ with SHARP requires an accurate characterisation of their magnitudes, sizes, and structural properties. To this end, we make use of the publicly released \textsc{COSMOS-Web} galaxy catalogue, whose construction and validation are detailed in \citet{Shuntov2025} and \citet{Huertascompany2025}. The catalogue is based on the COSMOS-Web JWST Cycle~1 Treasury Program ($\#$1727), a 255-hour survey that provides NIRCam coverage over $\sim0.54\,\mathrm{deg}^{2}$ in four filters (F115W, F150W, F277W, F444W), complemented by MIRI/F770W imaging and extensive legacy multi-wavelength observations from \textit{HST}, Subaru/HSC, UltraVISTA, CFHT, and \textit{Spitzer}. Compared to previous COSMOS releases, the new COSMOS2025 compilation represents a major upgrade: all photometry, structural parameters, redshifts, and derived physical quantities have been recomputed homogeneously starting from the JWST/NIRCam mosaics.

In this work, we adopt the stellar masses $M_{\star}$, photometric redshifts $z_{\rm phot}$, and specific star-formation rates (sSFR) provided by the \textsc{COSMOS-Web} catalogue. These quantities correspond to the median of the posterior probability distribution functions obtained through SED fitting with \textsc{LePhare}, using BC03 stellar population models, a range of exponentially declining and delayed star-formation histories, and multiple dust-attenuation prescriptions.

Likewise, the effective radii ($R_{\mathrm{e}}$) are taken directly from the \textsc{COSMOS-Web} catalogue. These were derived from the joint multi-band Sérsic fits applied to the NIRCam images (F115W–F444W). The JWST mosaics are sampled at $0.03''$/pixel, and the catalogue’s PSF modelling ensures reliable recovery of sub-kiloparsec radii out to $z\sim 4$ \citep{Shuntov2025}. The catalogue is approximately $90\%$ complete at $\log(M_{\star}/M_{\odot})\sim 9$ at $z\sim 6$, corresponding to an improvement of nearly 1\,dex in mass completeness relative to COSMOS2020 \citep{Weaver2022}.

Starting from the full catalogue, and following the prescription described in it, we construct a clean galaxy sample by selecting objects that satisfy:
\begin{itemize}
\item \texttt{LePHARE$_{type}$ = 0}, to include only galaxy sources;
\item \texttt{warn$_{flag}$ = 0}, to exclude objects whose ground-based photometry is contaminated by bright stars;
\item $\texttt{mag$\_$model$\_$F444W} < 30$, ensuring sufficient photometric quality. Here, $\texttt{mag$\_$model$\_$F444W}$ is the total F444W AB magnitude derived from the Sérsic model fits.
\end{itemize}

For the purposes of this feasibility study, we focus on massive passive galaxies in the redshift range $1.5 < z < 3$. This choice ensures that all key rest-frame optical absorption features required for robust stellar–population analysis, such as the higher–order Balmer lines, the Fe–sensitive indices up to Fe5790\,Å, and the Mgb triplet near 5175\,Å (see Fig. 1 in Longhetti et al. in this Science Book), remain fully covered within the wavelength range accessible to SHARP. At $z \sim 3$, these diagnostics are redshifted to $\lambda \approx 1.5$–2.4\,µm, i.e. entirely within the combined $H$– and $K$–band domain of VESPER.

Access to the $K$ band is a distinctive capability of SHARP: upcoming ELT instruments such as MOSAIC, even with its planned multi-IFU mode, will operate only up to the $H$ band, restricting detailed stellar-population studies to $z \lesssim 2.0$. By extending spectroscopic coverage to $\lambda \approx 2.4$\,µm, SHARP uniquely enables the measurement of age, metallicity, and abundance-ratio diagnostics all the way to $z \simeq 3$, and potentially beyond, making this redshift range the natural upper limit for our feasibility analysis.

Massive quiescent systems are identified as those with $\log(M_{\star}/M_{\odot}) > 10.5$ and $\log(\mathrm{sSFR/yr^{-1}}) < -10$, complemented by a rest-frame colour selection following the NUV–$r$–J criterion by \citet{Ilbert2013}:
\begin{equation}
    M_{\rm NUV} - M_R > 3\,(M_R - M_J) \quad \mathrm{and} \quad
    M_{\rm NUV} - M_R > 3.1.
\end{equation}
This combined selection yields a robust population of quiescent galaxies at  $z>1.5$ \citep{Huertascompany2025}. 

\begin{figure*}
    \centering
    \includegraphics[width=1\linewidth]{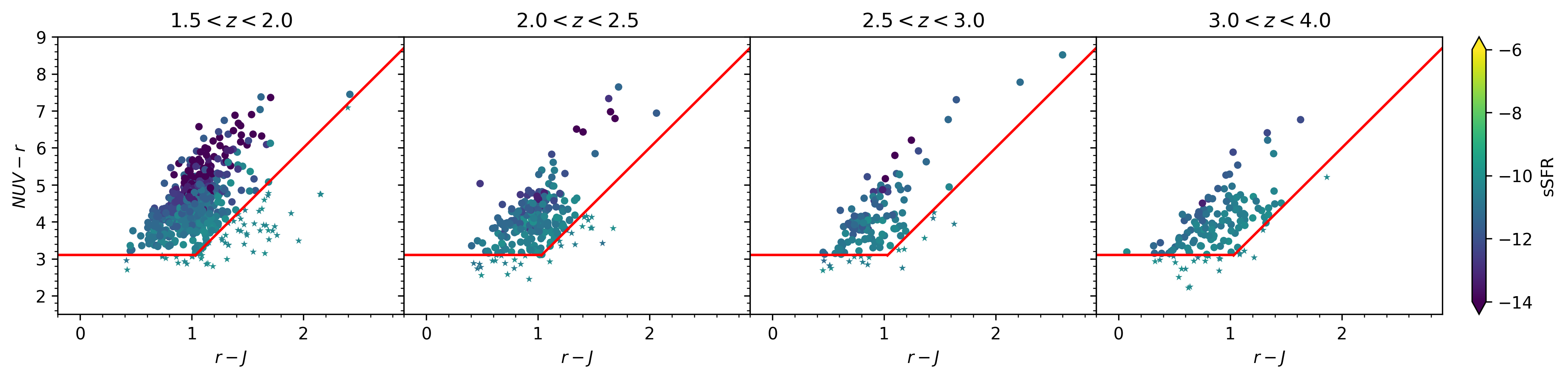}
    \caption{NUV–$r$–J colour–colour diagram for the massive passive sample in four redshift bins. Circle and star points show all galaxies with $\log(M_{\star}/M_{\odot})>10.5$ and $\log(\mathrm{sSFR/yr^{-1}})<-10$. Red lines indicate the passive selection criteria of \citet{Ilbert2013}. Circle symbols highlight galaxies satisfying both the sSFR and colour-based selections. The colour scale traces the sSFR.}
    \label{colorfig}
\end{figure*}

Figure~\ref{colorfig} shows the NUV–$r$–J colour–colour diagram for the selected galaxies in three redshift bins between $1.5$ and $3.0$, namely $1.5 \leq z < 2$, $2.0 \leq z < 2.5$, and $2.5 \leq z < 3$. For completeness, a fourth panel displays the colour–colour distribution of galaxies at higher redshift ($3.0 \leq z < 4.0$). Although this latter subsample is not included in the main target of our feasibility analysis, we show it because our final discussion illustrates that SHARP can extend the study of spatially resolved stellar population analysis to this regime, albeit with a reduced level of completeness compared to the core redshift range. All galaxies satisfying $\log(M_{\star}/M_{\odot})>10.5$ and $\log(\mathrm{sSFR/yr^{-1}})<-10$ are displayed; the red lines mark the \citet{Ilbert2013} colour selection, and circle symbols indicate galaxies that meet both the sSFR and colour-defined passive criteria. The colour scale encodes the sSFR.

\section{Observational Feasibility with SHARP/VESPER}\label{sec:feasibility}

The \textsc{COSMOS-Web} passive sample defined in Sect.~\ref{sec:sample} allows us to define the distribution of magnitudes and sizes of quiescent galaxies at $1.5<z<4$, which we use to assess the feasibility of measuring stellar population gradients with SHARP/VESPER.

\begin{figure*}
    \centering
    \includegraphics[width=1\linewidth]{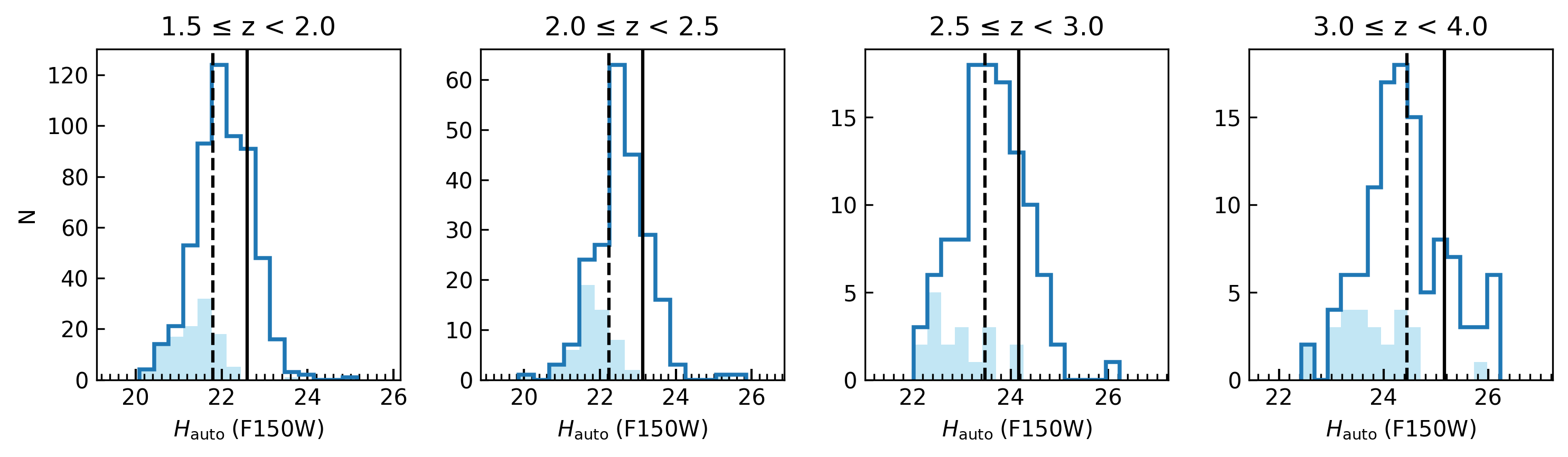}
    \caption{F150W magnitude distribution of the massive passive sample in four redshift bins (solid blue lines). The pale-blue shaded histogram shows the distribution of galaxies with $\log(M_{\star}/M_{\odot})>11$. In each panel, solid(dashed) line indicates the 80th-pecentile of the distribution for the full(most massive) sample.}
    \label{mag}
\end{figure*}

\begin{figure*}
    \centering
    \includegraphics[width=1\linewidth]{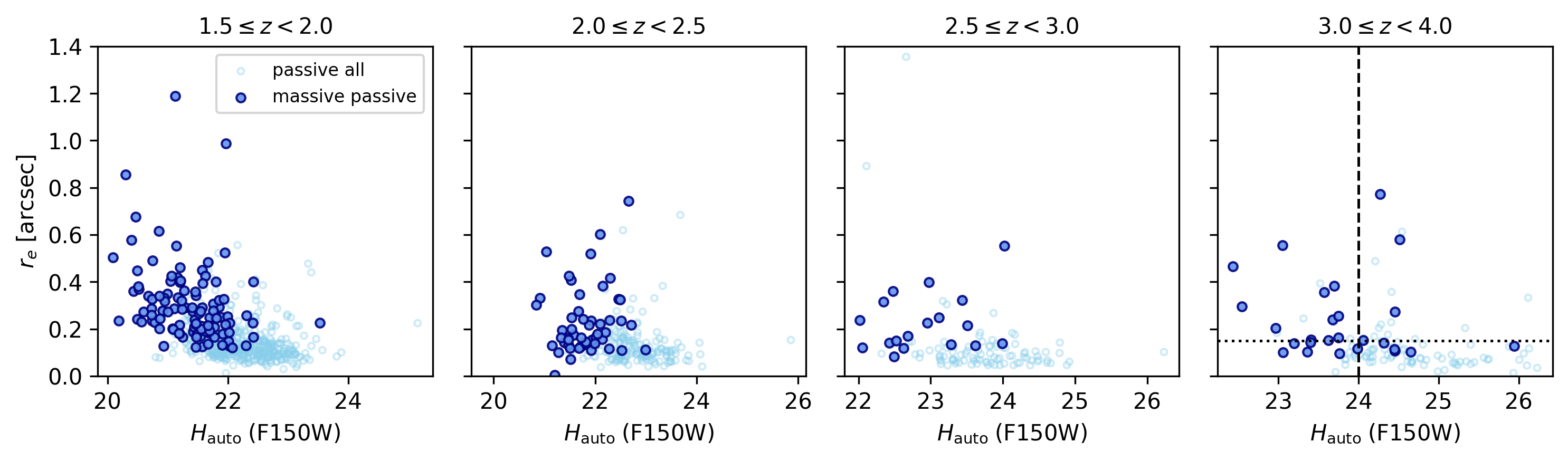}
    \caption{Effective radius versus F150W magnitude in four redshift bins. Pale-blue open circles show the full massive passive sample; filled symbols highlight galaxies with $\log(M_{\star}/M_{\odot})>11$. The dashed lines in the $3 \leq z < 4$ panel mark an illustrative region (H < 24 and $R_{\mathrm{e}} < 0.15''$) corresponding to one of the size–magnitude combinations for which SHARP is expected to reach $S/N$ levels suitable for measuring stellar population gradients within $\sim 30$ h even at this redshift (see Sec. 3.1. }
    \label{size}
\end{figure*}

Figure~\ref{mag} presents the F150W magnitude distribution of the massive passive sample in four redshift bins. The pale-blue shaded histogram highlights the subsample of galaxies with $\log(M_{\star}/M_{\odot})>11$. Figure~\ref{size} shows the corresponding distribution of effective radii $R_{\mathrm{e}}$ in arcseconds as a function of F150W magnitude in the same redshift bins; filled symbols mark the most massive systems.

From Figures~\ref{mag} and \ref{size}, we find that $\sim 80\%$ of the passive sample has $H \leq 24$ up to $z\sim 3$. 
For the most massive systems ($\log M_{\star}/M_{\odot}>11$), these thresholds become $H \leq 23.5$.
Table~\ref{mag_sum} summarises the 20th and 80th percentiles of the $R_{\mathrm{e}}$ distribution, together with the 80th-percentile magnitude in each redshift bin, for both the full sample and the most massive one.

\begin{table}
    \centering
    \caption{20th and 80th percentiles of the effective radius distribution, and 80th-percentile F150W magnitude ($H_{80}$), for the total passive sample and for galaxies with $\log(M_{\star}/M_{\odot})>11$, in three redshift bins.}
    \begin{tabular}{lccccc}
    \hline
    Redshift bin & $R_{\mathrm{e},20}$ & $R_{\mathrm{e},20}$ & $R_{\mathrm{e},80}$ & $R_{\mathrm{e},80}$ & $H_{80}$ \\
                 & (arcsec)            & (kpc)            &   (arcsec) & (kpc)      \\
    \hline
    \multicolumn{6}{c}{Total sample} \\
    \hline
    $1.5 \leq z < 2.0$  & 0.09 & 0.8 & 0.24 & 2,0 & 22.6 \\
    $2.0 \leq z < 2.5$  & 0.08 & 0.7 & 0.19 & 1.6 & 23.1 \\
    $2.5 \leq z < 3.0$  & 0.06 & 0.5 & 0.15 & 1.2 &  24.1 \\
    \hline
    \multicolumn{6}{c}{Massive sample ($\log M_{\star}/M_{\odot}>11$)} \\
    \hline
    $1.5 \leq z < 2.0$  & 0.19 & 1.6 & 0.40 & 3.4 & 21.7 \\
    $2.0 \leq z < 2.5$  & 0.13 & 1.1 & 0.34 & 2.8 & 22.2 \\
    $2.5 \leq z < 3.0$  & 0.13 & 1.0 & 0.32 & 2.6 & 23.5 \\
    \hline
    \end{tabular}
    \label{mag_sum}
\end{table}

\subsection{Exposure time estimates with the SHARP ETC}\label{sec:etc}

Since the VESPER spectral resolution element is approximately 5Å in the H-band, detailed stellar–population analyses can be performed with continuum signal-to-noise ratios of order 
$S/N \sim 10-15$ per resolution element. \citep[e.g.][]{Gallazzi2005,Saracco2020}. In this section, we use the official SHARP exposure time calculator (ETC v0.6) to quantify the exposure times (t$_{exp}$) needed to reach these $S/N$ levels for representative passive galaxies drawn from the \textsc{COSMOS-Web} sample.

For each redshift bin defined in Sect.~\ref{sec:feasibility}, we simulate two galaxies: a compact system with $R_{\mathrm{e}} = R_{\mathrm{e},20}$ and an extended one with $R_{\mathrm{e}} = R_{\mathrm{e},80}$ (Table~\ref{mag_sum}). In both cases, we adopt the 80th-percentile F150W magnitude $H_{80}$ as representative of the faint end of the massive passive population. This choice ensures that the resulting exposure times correspond to a conservative, ``worst-case'' scenario: if observations are feasible at $H_{80}$, they will be feasible for all brighter sources at the same redshift. Throughout this section, the terms ``compact'' and ``extended'' refer exclusively to the percentile-based size selection within our high-redshift sample; the extended systems should not be interpreted as analogues of local massive early-types of similar mass.

\begin{figure}
    \centering
    \includegraphics[width=1\linewidth]{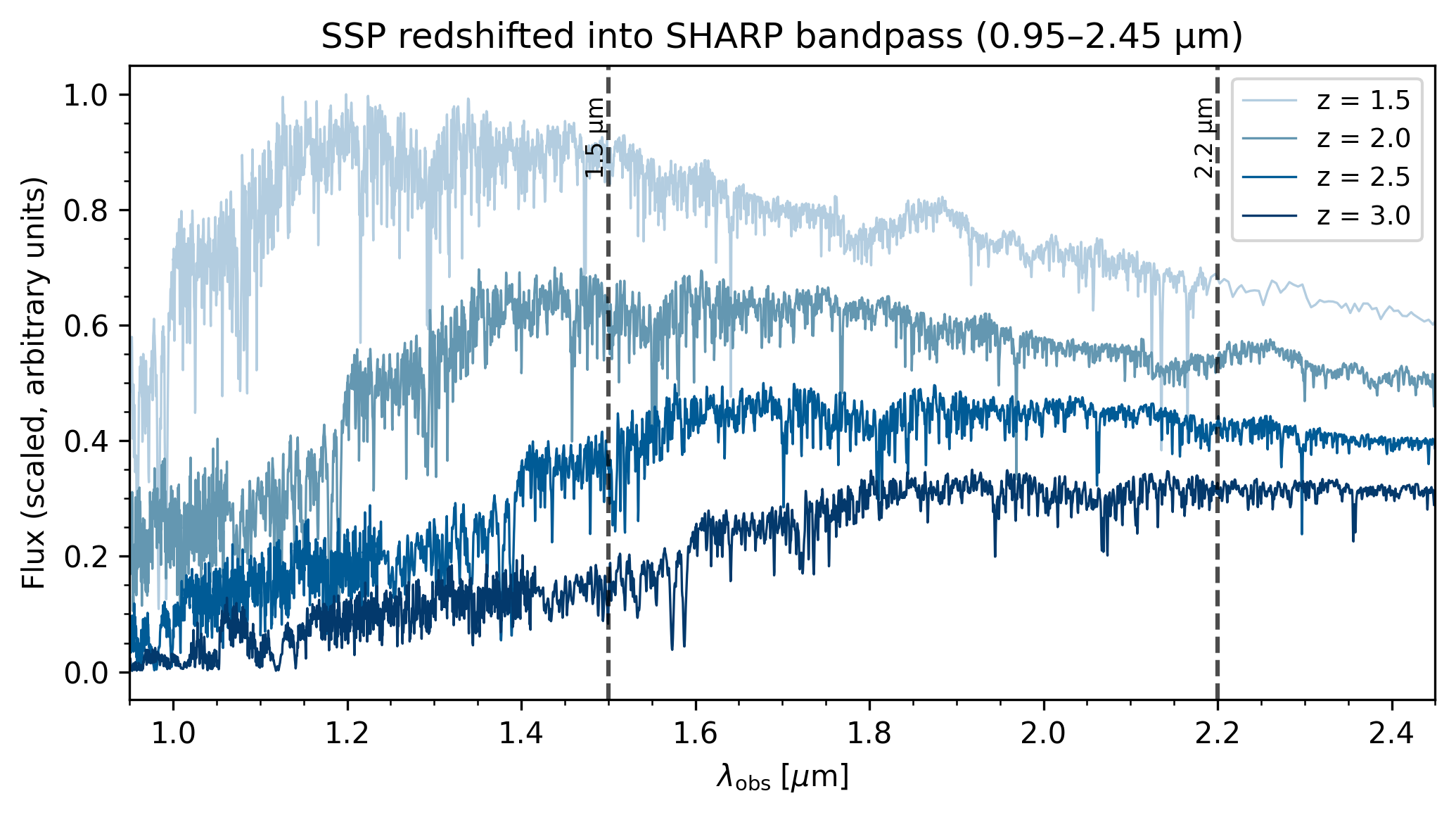}
    \caption{Redshifted spectra of a 2\,Gyr single stellar population template (solar metallicity, Bruzual $\&$ Charlot, 2016) shown at
$z=1.5$, $z=2$, $z=2.5$, and $z=3$ (from top to bottom). The strong continuum slopes characteristic of passive galaxies at these epochs are clearly visible. The vertical dashed lines mark the two reference wavelengths, $\lambda = 1.5\,\mu$m and $\lambda = 2.2\,\mu$m, at which the $S/N$ estimates discussed in Sect.~\ref{sec:etc} are evaluated.}
    \label{ssp}
\end{figure}

Because passive galaxies at $z>1.5$ display pronounced slopes in the optical rest frame continuum (Fig.~\ref{ssp}), we evaluate the required exposure times at two reference wavelengths, $\lambda = 1.5,\mu$m and $\lambda = 2.2,\mu$m, corresponding to the centres of the SHARP $H$- and $K$-band windows. For each simulated galaxy, we compute the exposure time needed to reach $S/N=10$ and $S/N=15$ at both wavelengths.

This dual–wavelength, dual–S/N strategy provides a more complete view of the expected data quality across the full spectral range: for any given exposure time, one can readily infer the approximate S/N achievable in both the $H$ and $K$ bands. For example, if a 20-hour exposure delivers $S/N\sim 15$ in $H$ out to $2R_{\mathrm{e}}$, the corresponding entries for $S/N=10$ and $S/N=15$ in the $K$ band indicate whether the same radial extent is attainable at longer wavelengths. In this way, the tabulated values allow observers to assess at a glance how performance in one band translates to the other and to plan observing strategies accordingly.

All ETC simulations assume a simple stellar population (SSP) of age 2\,Gyr and solar metallicity. We adopt an airmass of 1.5 and the default SHARP/VESPER instrumental and sky background configurations.

Motivated by recent morphological studies indicating that many high-redshift passive galaxies exhibit disk-like or moderately concentrated structures \citep[e.g.][]{Suess2019,Suess2022}, each mock galaxy is modelled with a Sérsic surface-brightness profile of index $n=3$. 
For every configuration, we extract spectra within three circular annuli:
$0 < R < 0.5R_{\mathrm e}$,
$0.5R_{\mathrm e} < R < R_{\mathrm e}$, and
$R_{\mathrm e} < R < 2R_{\mathrm e}$,
and we use the ETC to compute the exposure time needed to reach the target $S/N$ in each region. In this way, the simulations provide the integration times required to recover meaningful stellar population constraints not only in galaxy centres, but also into their outskirts.

For the most compact galaxies in our sample (those with $R_{\mathrm e} = R_{\mathrm e,20}$), the innermost aperture ($0.5R_{\mathrm e}$) corresponds to angular scales of approximately $(0.045,\,0.040,\,0.030)$ arcsec in the three redshift bins considered (see Tab.\,1). These scales are fully accessible with SHARP thanks to its spatial sampling of 30\,mas and to the diffraction-limited point-spread function delivered by MORFEO’s MCAO system (FWHM $\sim$ 15\,mas in $K$). In physical units, these radii correspond to roughly $(0.4,\,0.35,\,0.25)$\,kpc, demonstrating that SHARP will be able to probe the central regions of high-redshift passive galaxies at genuinely sub-kiloparsec resolution.

Table~\ref{tab:texp} summarises the exposure times (in hours) required to reach $S/N = 10$ and $S/N = 15$ at 1.5,$\mu$m (H–10, H–15) and 2.2,$\mu$m (K–10, K–15) for both compact and extended sources, across all redshift bins and radial annuli. Colour coding is used to guide the interpretation: values in green correspond to exposure times shorter than 10 hours, those in blue indicate 10–20 hours, and those in magenta mark 20–30 hours. We adopt 30 hours as a practical upper limit for feasible SHARP observations.

An inspection of Table~\ref{tab:texp} reveals several key trends.
For compact galaxies, exposure times of approximately 20\,h are sufficient to recover stellar population gradients out to $2R_{\mathrm{e}}$ down to the faint end of the passive population for all systems at $z<2.5$, and even shorter integrations ($\sim10$\,h) are adequate for the most massive galaxies. In the last bin range $2.5 < z < 3$, gradients remain measurable out to $\sim R_{\mathrm{e}}$ with exposure times of $\sim30$\,h.

For extended galaxies, 20\,h exposures enable robust measurements out to $R_{\mathrm{e}}$ for the total sample at $z<2.5$, and out to $2R_{\mathrm{e}}$ for the most massive systems. Reaching $2R_{\mathrm{e}}$ for the full population generally requires $\sim30$\,h.
In the highest redshift bin ($2.5<z<3$), the exposure times required to achieve adequate signal to noise ratios for extended galaxies with $H = H_{80}$ become prohibitive, making gradient measurements infeasible within $t_{\rm exp}\leq 30$ h. Nevertheless, the ETC results show that a significant fraction of the population remains accessible. In particular, for the total(most massive) sample, galaxies with magnitudes brighter than the median, that is $H < H_{50}=23.5$($H_{50}=22.7$), can reach $S/N=15$ in H-band in circular annuli out to $1R_{\mathrm{e}}$ within $t_{\rm exp}\lesssim 30$\,h.

A similar conclusion applies to the highest-redshift bin ($3\leq z<4$). Although the population as a whole is too faint to allow gradient measurements within reasonable exposure times, a subset of bright and compact systems remains within reach.
As an illustrative example (one of several viable size–magnitude combinations explored with the ETC), at $3\leq z<4$, in $\sim30$\,h it is possible to reach $S/N=10$ in the $H$ band and $S/N=15$\footnote{In these simulations we have adopted a SSP of 1\,Gyr.} in the $K$ band in circular annuli out to $1R_{\mathrm{e}}$ for galaxies with $H<24$ and $R_{\mathrm{e}}\simeq 0.15''$ (see Fig.\,3).
These examples are not exhaustive but demonstrate that, beyond the global trends captured in Table~\ref{tab:texp}, a meaningful subset of the high-redshift passive population remains fully accessible to SHARP. 

\begin{figure}
    \centering
    \includegraphics[width=1\linewidth]{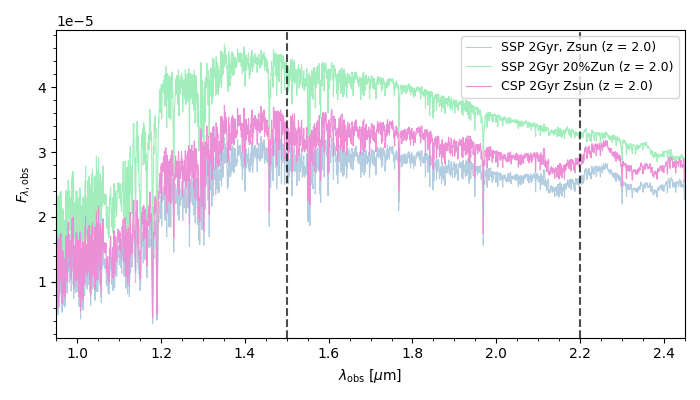}
    \caption{Comparison of different stellar population models used to test the robustness of our ETC results. The blue curve shows the reference simple stellar population (SSP) with age 2 Gyr and solar metallicity adopted throughout this work. The green curve represents an SSP with the same age but lower metallicity (20$\%$ Z$_{\odot}$), while the magenta curve shows a composite stellar population (CSP) with an exponentially declining star formation history ($\tau = 0.2$ Gyr), solar metallicity, and age of 2 Gyr. The dashed vertical lines indicate the wavelength ranges explored with the ETC simulations.}
    \label{ssp_comparison}
\end{figure}

All the conclusions presented are based on an SSP with an age of 2 Gyr and solar metallicity. We have verified that our results are robust against reasonable variations in these assumptions. In Fig.~\ref{ssp_comparison}
we compare the adopted SSP with alternative models, including a lower-metallicity SSP (20\% Z$_{\odot}$) at the same age, and a composite stellar population (CSP) with an exponentially declining star formation history ($\tau = 0.2$ Gyr), with age of 2 Gyr and solar metallicity. 
As shown in the figure, the continuum level within the wavelength ranges explored with the ETC is comparable to, or higher than, that of the adopted SSP. This ensures that our conclusions can be safely extended to more general and realistic stellar population models.

These trends confirm that SHARP/VESPER will be capable of mapping stellar population gradients for a representative range of passive galaxies out to $z \sim 3$ and even beyond, with exposure times well within standard observing allocations. Importantly, all the results shown in Tab.~\ref{tab:texp} were obtained adopting the 80th-percentile (i.e. comparatively faint) magnitudes in each redshift bin. As a consequence, the exposure times reported represent conservative upper limits: for the majority of the sample the achievable signal-to-noise ratios will be higher, and consequently robust gradient measurements will require substantially shorter integrations. The simulations presented in this work therefore provide a realistic and secure foundation for planning SHARP observations aimed at constraining the early assembly histories of passive galaxies.

\section{The Big Step Forward}\label{sec:bigstep}

The results presented in this work demonstrate that SHARP/VESPER will deliver a transformative advance in our ability to measure spatially resolved stellar population gradients in passive galaxies at $z>1.5$. To appreciate the magnitude of this improvement, it is instructive to compare the expected capabilities of SHARP with the current state of the art achieved with JWST.

The deepest resolved stellar population measurements for high-redshift quiescent galaxies to date come from the JWST–SUSPENSE program \citep{Cheng2025}. Using ultra-deep NIRSpec–MSA observations, Cheng et al.\ analysed a sample of eight passive galaxies at $1.1<z<2.1$, typically with total integration times of $\sim$16–20 hours. Thanks to the multiplexing capabilities of the MSA, the authors were able to extract spectra in two coarse spatial regions: an inner aperture covering approximately $R \lesssim 0.5R_{\mathrm e}$, and an outer aperture extending, on average, to $R \sim 1.5R_{\mathrm e}$. The innermost aperture used for the analysis corresponds to angular scales of roughly $0.1$–$0.15''$, which translate to physical radii of order $\sim 1$\,kpc at the redshifts of the SUSPENSE galaxies. This matches the intrinsic spatial resolution limit of JWST/NIRSpec, whose sampling of approximately $0.1''$ sets the maximum level of spatial detail achievable with MSA observations. Although these measurements represent a major breakthrough, providing the first direct constraints on stellar population gradients at these epochs, the spatial sampling remains necessarily sparse.

Pushing the resolution significantly below the $1 \text{ kpc}$ scale is not merely a quantitative improvement, but a physical necessity for several reasons. First, coarse sampling inevitably dilutes and smears out stellar population gradients, potentially washing out mild radial variations that are crucial for identifying slow evolutionary paths. Furthermore, as shown in Table 1, the typical sizes of dense, massive galaxies at $z > 2$ are of the order of $\sim 1 \text{ kpc}$; for these compact systems, the current JWST resolution limit effectively collapses most of the galaxy into a single resolution element, making any internal spatially resolved analysis impossible. Most importantly, empirical evidence suggests that the physical processes occurring within this central kiloparsec are the fundamental drivers behind the quenching of the entire system \citep[e.g.][]{Whitaker2017, Saracco2017}.
Resolving sub-kpc scales is therefore essential to distinguish between a 'slow' inside-out quenching—manifesting as a smooth, continuous age gradient \citep[e.g.][]{Tacchella16} and 'fast' transformative events, such as AGN-driven blowouts or nuclear starbursts. While the latter would leave a distinct 'core-envelope' discontinuity, JWST observations inevitably smear this signature into a single spatial average. Accessing scales $< 0.5 \text{ kpc}$ thus provides the only direct way to test the 'compaction' scenario \citep[][]{Dekel14, Zolotov15} and determine whether the bulge grew through steady star formation or via a rapid, dissipative event followed by central dynamical stabilisation.
Leveraging exposure times comparable to those used in SUSPENSE, SHARP/VESPER is poised to bridge this gap and catalyze the next decisive leap in extragalactic astrophysics. VESPER’s $30 \text{ mas}$ spatial sampling, combined with the diffraction-limited PSF delivered by MORFEO, will provide an angular resolution three times superior to JWST, finally unlocking the sub-kpc scales where bulge growth and the physics of quenching actually originates. 
By resolving these innermost regions, SHARP will shift the debate from when galaxies quenched to the specific physical mechanism, be it central dynamical stabilisation (morphological quenching) or feedback-driven gas expulsion—that governs the lifecycle of massive galaxies at their peak epoch of formation.

SHARP/VESPER is uniquely positioned to address this open question: its combination of high spatial resolution, broad spectral coverage, and integral-field capability will allow, for the first time, the spatial mapping of absorption-line diagnostics out to $2R_{\mathrm e}$ in typical quiescent galaxies at $z \sim 2$–$3$. This represents a significant breakthrough in our ability to disentangle fast, dissipative formation pathways from more extended or accretion-driven modes of assembly.

A final consideration concerns the number density of high-redshift quiescent galaxies. As noted by Longhetti et al. in this Science Book, their surface density is too low for passive-galaxy programs to take full advantage of the multiplexing capabilities of VESPER on their own (12 integral field unit over an area of 24" $\times$ 70"). However, the same authors emphasise that quiescent galaxies can be efficiently combined with several complementary science cases that share the same observing conditions and instrument configuration, such as star-forming galaxies, AGN, or proto-cluster members, within a single SHARP pointing. This strategy maximises the scientific return of each observation while still enabling meaningful studies of high-redshift passive systems.

Crucially, this multiplexing capability places SHARP at a clear advantage compared to future ELT instruments such as HARMONI. Although HARMONI offers comparable wavelength coverage (including the $K$ band, unlike MOSAIC), it operates with a single IFU, restricting each exposure to one primary target at a time. JWST/NIRSpec IFU is subject to the same fundamental limitation: it observes only one source per pointing and, in addition, its spatial resolution is set by a $\sim 0.1''$ sampling. By contrast, SHARP can observe multiple galaxies simultaneously, whether or not they belong to the same science programme, while delivering far superior spatial resolution, dramatically increasing both efficiency and scientific reach. 

In conclusion, SHARP/VESPER will not merely extend the reach of existing JWST studies: it will enable a fundamentally new regime of spatially resolved stellar population analysis at $z>2$, opening a window onto the physical processes that shaped the earliest generations of massive passive galaxies.

\section{Summary}

Recent NIRCam and NIRSpec observations have shown that massive quiescent galaxies ($\log(M_{\star}/M_{\odot})>10.5$) are already in place at $z>1.5$–2, exhibiting a wide diversity in size, stellar mass density, and structural properties. This heterogeneity suggests that multiple evolutionary pathways contributed to their early assembly, producing both compact and extended systems. Yet the relative importance of rapid dissipative events, inside-out growth, and the accretion of external material remains poorly constrained.

To understand how these galaxies formed, spatially resolved measurements are crucial, because radial variations in their stellar populations retain the imprint of the physical processes that governed their mass assembly and quenching.

At high redshift, rest-frame colour gradients have provided early evidence for the existence of stellar population gradients and their connection to galaxy structure. Yet photometry alone cannot break the degeneracy between age, metallicity, and dust attenuation, and therefore cannot reveal the mechanisms responsible for quenching and mass assembly. Spectroscopic measurements are needed, but current facilities lack the sensitivity, wavelength coverage, and spatial resolution required to obtain them beyond $z \sim 1.5$.

SHARP, the high-resolution near-infrared spectrograph planned for the ELT, together with its VESPER integral-field unit, is designed to overcome these limitations. The collecting area of the 39\,m aperture, combined with MORFEO’s adaptive optics system, will deliver the sensitivity and spatial resolution needed to probe the rest-frame optical spectra of passive galaxies at high redshift. VESPER will provide 30\,mas spatial sampling and diffraction-limited performance across the full 0.95–2.45\,µm range, giving access to all the absorption features required to constrain stellar ages, metallicities, and abundance ratios at $1.5 < z < 3$. 

In this work, we present a feasibility study that quantifies SHARP’s ability to measure stellar population gradients in massive passive galaxies with $\log(M_{\star}/M_{\odot})>10.5$ over $1.5<z<3$, and to explore under which conditions observations may be extended to even higher redshift. Using realistic massive passive galaxy populations from \textsc{COSMOS-Web} and detailed simulations with the official SHARP exposure-time calculator, we estimate the exposure times required to recover radial trends in age, metallicity, and abundance ratios for both compact and extended systems.

Our simulations show that, with exposure times comparable to those routinely used on 8–10,m class telescopes, SHARP will recover stellar population gradients out to roughly two effective radii for the majority of massive passive galaxies at $z<2.5$, and will still reach at least one effective radius up to $z\simeq3$. Crucially, the combination of VESPER’s 30,mas spatial sampling and MORFEO’s diffraction-limited performance will grant, for the first time, direct spectroscopic access to the inner kiloparsec of high-redshift quiescent galaxies. This central region is widely recognised as the engine room of galaxy evolution: it is where bulges assemble, where central mass density rises most rapidly, and where the physical conditions leading to quenching are established. Being able to resolve and characterise this scale marks a decisive leap forward, enabling us to probe the processes that set galaxies onto their passive evolutionary path at the epoch when these transformations first occurred (see Mancini et al. in this Science Book).

By enabling continuous radial sampling of passive galaxies at $z\sim2$–3, SHARP will deliver the first definitive view of the spatially resolved processes that governed their early assembly. This represents a major step forward in linking the physical mechanisms active during the peak of galaxy formation to the properties of massive quiescent systems in the local Universe, firmly establishing SHARP as a transformative instrument for high-redshift galaxy evolution studies.

\section*{Acknowledgments}
The SHARP team acknowledges support by Bando Ricerca Fondamentale INAF
2022,
Techno-Grant "SHARP" - 1.05.12.02.01 and Bando Ricerca Fondamentale INAF
2024,
Large-Grant "SHARP" - 1.05.24.01.01.

\begin{landscape}
\begin{table}
\centering
\caption{Exposure times (in hours) required to reach $S/N=10$ and $S/N=15$ per spectral resolution element at 1.5\,$\mu$m (H–10, H–15) and 2.2\,$\mu$m (K–10, K–15), for compact ($R_{\mathrm{e},20}$) and extended ($R_{\mathrm{e},80}$) sources in each redshift bin. Green, blue, and magenta colors highlights t$_{exp}$$\leq10$, $10<$t$_{exp}$$\leq20$ and $20<$t$_{exp}$$\leq30$. Missing numerbs indicate t$_{exp}$$>100$ hours}
\label{tab:texp}

\begin{adjustbox}{max width=\linewidth,center}
\begin{tabular}{lccccccc}
\toprule
        & & \multicolumn{3}{c}{Compact sources ($R_{\mathrm{e},20}$)} & \multicolumn{3}{c}{Extended sources ($R_{\mathrm{e},80}$)} \\
        \cmidrule(lr){3-5} \cmidrule(lr){6-8}
    Redshift bin & S/N target &
    $t_{\rm exp}(0\!<\!R\!<\!0.5R_{\mathrm{e}})$ &
    $t_{\rm exp}(0.5R_{\mathrm{e}}\!<\!R\!<\!R_{\mathrm{e}})$ &
    $t_{\rm exp}(R_{\mathrm{e}}\!<\!R\!<\!2R_{\mathrm{e}})$ &
    $t_{\rm exp}(0\!<\!R\!<\!0.5R_{\mathrm{e}})$ &
    $t_{\rm exp}(0.5R_{\mathrm{e}}\!<\!R\!<\!R_{\mathrm{e}})$ &
    $t_{\rm exp}(R_{\mathrm{e}}\!<\!R\!<\!2R_{\mathrm{e}})$ \\
    \midrule
    
    \multicolumn{8}{c}{\textbf{Total sample}} \\
    \midrule
    $1.5 \leq z < 2.0$ & H–10 & $\bluegreen{0.6}$ & $\bluegreen{2.8}$ & $\bluegreen{3.6}$ & $\bluegreen{2.4}$ & $\bluegreen{7.3}$  & $\bluebold{13.8}$ \\
                       & H–15 & $\bluegreen{1.4}$ & $\bluegreen{6.3}$ & $\bluegreen{8.3}$ & $\bluegreen{5.5}$ & $\bluebold{16.4}$ & $\bluemag{30}$ \\
                       & K–10 & $\bluegreen{1.1}$ & $\bluegreen{5.3}$ & $\bluegreen{7.1}$ & $\bluegreen{4.6}$ & $\bluebold{14.1}$ & $\bluemag{26.7}$ \\
                       & K–15 & $\bluegreen{2.5}$ & $\bluebold{12.2}$ & $\bluebold{15.9}$ & $\bluebold{10.7}$ & $\bluemag{31.7}$ & 60.1 \\
    \hline
    $2.0 \leq z < 2.5$ & H–10 & $\bluegreen{1.4}$ & $\bluegreen{3.5}$ & $\bluegreen{7.2}$ & $\bluegreen{3.6}$ & $\bluebold{11.2}$ & $\bluemag{21.1}$ \\
                       & H–15 & $\bluegreen{3.2}$ & $\bluegreen{7.9}$ & $\bluebold{17.3}$ & $\bluegreen{8.2}$ & $\bluemag{25.1}$ & 48.0 \\
                       & K–10 & $\bluegreen{2.6}$ & $\bluegreen{6.6}$ & $\bluebold{14.7}$ & $\bluegreen{7.0}$ & $\bluemag{21.6}$ & 41.0 \\
                       & K–15 & $\bluegreen{5.7}$ & $\bluebold{14.9}$ & 33.0 & $\bluebold{15.7}$ & 48.0 & 93.0 \\
    \hline
    $2.5 \leq z < 3.0$ & H–10 & $\bluegreen{4.4}$ & $\bluegreen{7.9}$ & $\bluemag{33.2}$ & $\bluemag{21.5}$ & 48.9 & -- \\
                       & H–15 & $\bluegreen{10.0}$ & $\bluebold{17.7}$ & 74.0 & 46.5 & --   & -- \\
                       & K–10 & $\bluegreen{3.6}$ & $\bluegreen{6.5}$ & $\bluemag{27.2}$ & $\bluebold{17.0}$ & 40.4 & -- \\
                       & K–15 & $\bluegreen{7.9}$ & $\bluebold{15.6}$ & 61.3 & 39.1 & 91.9 & -- \\
    \midrule

        \multicolumn{8}{c}{Massive sample ($\log M_{\star}/M_{\odot}>11$)} \\
        \midrule
    $1.5 \leq z < 2.0$ & H–10 & $\bluegreen{0.4}$ & $\bluegreen{1.1}$ & $\bluegreen{2.0}$ & $\bluegreen{0.9}$ & $\bluegreen{2.7}$  & $\bluegreen{6.4}$ \\
                       & H–15 & $\bluegreen{0.8}$ & $\bluegreen{2.0}$ & $\bluegreen{4.4}$ & $\bluegreen{2.0}$ & $\bluegreen{6.1}$ & $\bluebold{14.3}$ \\
                       & K–10 & $\bluegreen{0.6}$ & $\bluegreen{2.0}$ & $\bluegreen{3.8}$ & $\bluegreen{1.7}$ & $\bluegreen{5.3}$ & $\bluebold{12}$ \\
                       & K–15 & $\bluegreen{1.5}$ & $\bluegreen{4.4}$ & $\bluegreen{8.0}$ & $\bluegreen{3.9}$ & $\bluebold{11.8}$ & $\bluemag{27.9}$ \\
    \hline
    $2.0 \leq z < 2.5$ & H–10 & $\bluegreen{0.5}$ & $\bluegreen{1.1}$ & $\bluegreen{2.8}$ & $\bluegreen{1.6}$ & $\bluegreen{4.5}$ & $\bluebold{10.2}$  \\
                       & H–15 & $\bluegreen{1.0}$ & $\bluegreen{2.4}$ & $\bluegreen{6.4}$ & $\bluegreen{3.6}$ & $\bluegreen{10.0}$ & $\bluebold{23.0}$  \\
                       & K–10 & $\bluegreen{0.9}$ & $\bluegreen{2.0}$ & $\bluegreen{5.4}$ & $\bluegreen{3.0}$ & $\bluegreen{8.5}$ & $\bluebold{19.6}$  \\
                       & K–15 & $\bluegreen{2.0}$ & $\bluegreen{4.5}$ & $\bluebold{12.2}$ & $\bluegreen{6.8}$ & $\bluebold{19.0}$ & 44  \\
    \hline
    $2.5 \leq z < 3.0$ & H–10 & $\bluegreen{7.0}$ & $\bluebold{17.0}$ & 36.0 & $\bluebold{19.0}$ & 62.3 & -- \\
                       & H–15 & $\bluebold{15.9}$ & 40 & 81.0 & 43.2 & --   & -- \\
                       & K–10 & $\bluegreen{5.8}$ & $\bluebold{14.7}$ & $\bluemag{30.1}$ & $\bluebold{16.4}$ & 51.4 & -- \\
                       & K–15 & $\bluebold{13.0}$ & $\bluemag{30}$ & 67.2 & 36.3 & -- & -- \\
\bottomrule
\end{tabular}
\end{adjustbox}
    \label{tab:texp}
\end{table}
\end{landscape}

\clearpage 

\printcredits

\bibliographystyle{cas-model2-names}

\bibliography{cas-refs}



\end{document}